\documentclass[aps,prb,twocolumn,superscriptaddress,amsmath,amssymb,showpacs]{revtex4}
\usepackage{graphicx}
\usepackage{dcolumn}
\usepackage{bm}

\begin{document}

 \title{Doping, density of states and conductivity in polypyrrole and poly({\it p-}phenylene vinylene)}

 \author{I. N. Hulea}

 \affiliation{Kamerlingh Onnes Laboratory, Leiden University, POB. 9504, 2300 RA Leiden, The Netherlands}

 \affiliation{Dutch Polymer Institute (DPI), POB. 902, 5600 AX Eindhoven, The Netherlands}

 \author{H. B. Brom}

 \affiliation{Kamerlingh Onnes Laboratory, Leiden University, POB. 9504, 2300 RA Leiden, The Netherlands}

 \author{A. K. Mukherjee}

 \affiliation{Indian Institute of Science, Department of Physics, Bangalore, India}

 \author{R. Menon}

 \affiliation{Indian Institute of Science, Department of Physics, Bangalore, India}

\date{published in Phys. Rev. B {\bf 72} on August 19}

\begin{abstract}
The evolution of the density of states (DOS) and conductivity as
function of well controlled doping levels in
OC$_1$C$_{10}$-poly({\it p-}phenylene vinylene)
[OC$_1$C$_{10}$-PPV] doped by FeCl$_3$ and PF$_6$, and PF$_6$
doped polypyrrole (PPy-PF$_6$) have been investigated. At a doping
level as high as 0.2 holes per monomer, the former one remains
non-metallic, while the latter crosses the metal-insulator
transition. In both systems a similar almost linear increase in
DOS as function of charges per unit volume ($c^*$) has been
observed from the electrochemical gated transistor data. In
PPy-PF$_6$, when compared to doped OC$_1$C$_{10}$-PPV, the energy
states filled at low doping are closer to the vacuum level; by the
higher $c^*$ at high doping more energy states are available,
which apparently enables the conduction to change to metallic.
Although both systems on the insulating side show $\log \sigma
\propto T^{-1/4}$ as in variable range hopping, for highly doped
PPy-PF$_6$ the usual interpretation of the hopping parameters
leads to seemingly too high values for the density of states.
\end{abstract}

\pacs{72.80.Le, 71.20.Rv, 72.20.Ee, 73.61.Ph}

\maketitle

\section{Introduction}
Since the discovery of conducting polyacetylene (PAc) at the end
of the seventies,\cite{Chiang77} charge transport mechanisms in
semiconducting and conducting polymers have been of great
interest. In polypyrrole (PPy), as in PAc, a transition from
insulating (zero dc-conductance for temperature $T$ going to zero)
to metallic state (non-zero dc-conductance in the limit of zero
Kelvin) occurs by increasing the doping level;\cite{Martens01b}
and metallic PPy, among highly doped conducting polymers, is one
of the most widely studied due to its environmental stability,
which makes it attractive for technological applications. Usually
in conducting polymers, doping adds or removes electrons to the
$\pi$-band formed by the overlapping $p$-orbitals in the
conjugated polymer backbone. Although the electrons in the
$\pi$-band could be delocalized, not all conjugated polymers can
be brought into the metallic state. For example,
polyalkylthiophenes (PAT) and
poly[2-methoxy-5-($3'$,$7'$–dimethyloctyoxy)-$p$-phenylene
vinylene (OC$_1$C$_{10}$-PPV),\cite{notePPV} that have been
frequently used in polymeric transistors and polymeric light
emitting diodes, respectively, remain as insulators even at the
highest doping levels (with dopants like FeCl$_3$).
\cite{Reedijk99,Romijn01,Martens03}

To explain the transport data in conducting polymers in general,
key ingredients are the crystalline coherence length (a few
nanometers), the volume fraction of crystallinity ($> 50$\%), the
doping level, the interchain transfer integral, the energy
dependence of the density of states, the extent of disorder in the
material, charge repulsion and polaronic effects.
\cite{Kohlman97,Kaiser01,Prigodin04,Romijn03,Martens04} The
relevant values of the transfer integral, the spread in its mean
value due to disorder and of the Coulomb correlations are usually
all around 0.1 eV or less, which is close to the thermal energy at
300 K. A systematic study of the evolution of density of states
(DOS) and charge transport as a function of well controlled doping
level is still lacking in several conducting polymers. In this
work the difference between FeCl$_3$ and PF$_6$ doped
OC$_1$C$_{10}$-PPV and PF$_6$ doped PPy, as a function of doping
level is investigated in detail by studying both the
electrochemical gated transistor (EGT) characteristics and
temperature dependence of conductivity using a precise calibration
of the amounts of doping. The higher DOS per unit volume for
PF$_6$ doped PPy compared to doped OC$_1$C$_{10}$-PPV and the
occupation of the energy states near the Fermi level explain the
observed difference in conductivity behavior.

\section{Experiment}

OC$_1$C$_{10}$-PPV was doped in solution with iron(III)chloride,
FeCl$_3$. Ideally, the following redox reaction should take place:
$\rm PPV + 2FeCl_3 \rightarrow PPV^+ + FeCl_2 + FeCl_4^-$. Films
were obtained by slowly evaporating the solvent. \cite{Reedijk99}
Under ambient conditions, the conductive properties of the films
were stable over several weeks. Polypyrrole doped by PF$_6$
(PPy-PF$_6$) was polymerized and doped by anodic oxidation in an
electrochemical cell with glassy carbon electrode and platinum
foil as working and counter electrodes, respectively. The
polymerization was carried out at - $40^0$C under nitrogen
atmosphere to improve the structural order in the system, and the
samples were systematically dedoped to attain the desired doping
level.\cite{Yoon94,Lee95} Free-standing films (thickness $\sim$ 20
microns) were used for conductivity measurements; and the films on
glass substrate, on which Au-contacts were evaporated before
deposition, were used for electrochemical gated transistor (EGT)
experiments. In the EGT measurements on PPV and PPy the hole
charge was counterbalanced by PF$_6^{-}$ anions from the
electrolyte solution. \cite{Hulea04}

\section{Results and discussion}

\indent {\it Doping Level and Density of States}. The FeCl$_3$
doping levels in the PPV samples used for the $T$ dependence of
$\sigma$ were calculated from the amount of chemicals used in the
solutions, and further investigated in detail by Fe M\"{o}ssbauer
measurements.\cite{Martens03} The doping levels discussed in this
work are between 0.02 and 0.33 charges per monomer ($c$). Also,
earlier studies have shown that by using the semiconducting
polymer in an EGT $c$ can be obtained by summing the integrated
currents, which are directly measured as described below; $c$
ranges from 10$^{-4}$ up to 0.4. The PF$_6$ levels in the PPy
samples in the $T$ dependence of $\sigma$ data were deduced from
$^{19}$F-NMR and by using the sum rule for $\sigma(\omega)$ too
(for details see appendix A); $c$ lies between 0.065 and 0.23. In
PPy-EGT, the $c$-values discussed here range between $10^{-4}$ and
$10^{-2}$ charges per monomer; at higher doping levels the
measurements were not reversible and reproducible. Doping levels
can also be expressed per nm$^3$ ($c^*$), by knowing the estimated
volume of a monomer (ring volume), which in PPy is 0.13 nm$^3$ and
in PPV is 0.48 nm$^3$.\cite{Pouget94} Especially the latter
convention will be used in the following.

\begin{figure}[htb]
\begin{center}
 \includegraphics[width=8cm]{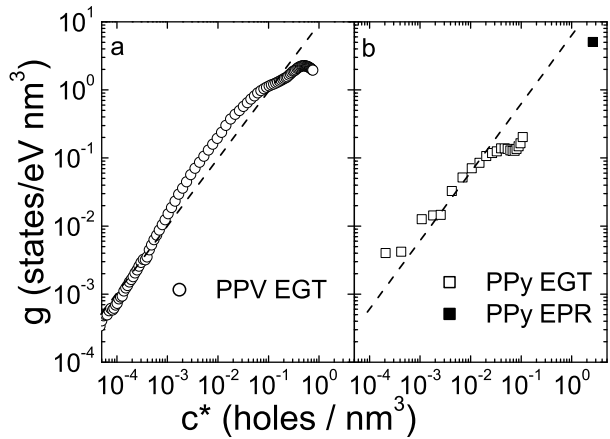}
\end{center}
\noindent{\caption{$g$ vs $c^*$ on a double log. scale. (a) EGT
data for PF$_6$ doped PPV. The found dependence is roughly linear
at low doping. The dashed line corresponds to $g \propto c^*$. (b)
PF$_6$ doped PPy from EGT and ESR (at $c^*$ =2.54 charges per
nm$^3$) data. \protect\cite{Joo00} The found dependence is again
almost linear (dashed line) at low doping.}\label{PPVgvsc}
\label{PPygvsc}}
\end{figure}

\begin{figure}[htb]
\begin{center}
\includegraphics[width=6cm]{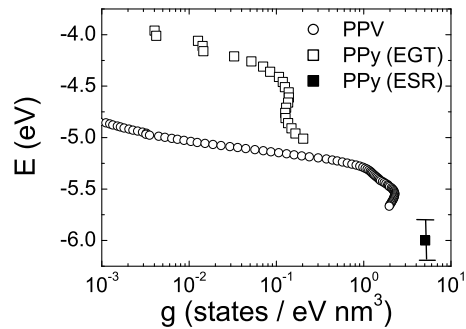}
\end{center}
\noindent{\caption{The $E$ dependence of the experimentally
determined DOS of PPy (per eV and nm$^3$) compared to PPV on a
linear-log. scale.\cite{Hulea04} Due to chemical instability of
PPy in the EGT no data points at high doping levels are
available.} \label{PPyEvsg}}
\end{figure}

The density of states will be expressed as the number of states
per eV per nm$^3$, and is denoted by $g$. In an EGT study $g$ is
determined as a function of energy. It equals the number of
elementary charges $\Delta Q/e$ that can be stored in the polymer
in a small step of the electrochemical potential ($\mu_e$) of
$\approx$ 10 meV, divided by the number of monomers and the
monomer volume. This number can be easily calculated. The
concentration at a given voltage is obtained via summation of all
$\Delta Q/e$ up to that value. The $g$ vs $c^*$ data for PF$_6$
doped PPV are shown in Fig.~\ref{PPVgvsc}a. The data follow a
linear dependence $g \propto c^*$ (the dashed line), especially at
lower doping levels. Because $dc^*/d\mu_e = g(\mu_e)$ it means
that $g \propto \exp(\mu_e)$ up to 0.5 states per eV per nm$^3$.
At higher values of $g(E)$ the dependence on $E$ becomes Gaussian,
as shown in previous work. \cite{Hulea04} The $g$ vs $c^*$ data
for PF$_6$ doped PPy are shown in Fig.~\ref{PPygvsc}b. The EGT
data are almost linear in $c^*$, and stable only at low PF$_6$
concentrations. To extrapolate the behavior of $g$ vs $c^*$, the
data point at high doping level $c^* = 2.54$, from an ESR study by
Joo {\it et al.},\cite{Joo00} is included. From the same ESR
study, the DOS at the Fermi level per spin was determined to be
0.33 states $/$ (eV monomer) for a metallic sample of PF$_6$ doped
PPy.

A comparison of $E$ vs $g$ in both systems is shown in
Fig.~\ref{PPyEvsg}. Knowing the Ag reference electrode location at
4.47 V below the vacuum level, the electrochemical potentials
could be correlated with the vacuum level.\cite{Hulea04} Based on
the EGT data with the additional data point from ESR, the tail of
the distribution of the hole states (at doping levels below 1\%)
in PPy-PF$_6$ is seen to be wider than in PPV, and also the
maximum in $g(E)$ is higher. However, a full comparison is
hindered by the absence of reliable PPy data from EGT above 0.2
states$/$nm$^3$eV.

\begin{figure}[htb]
\begin{center}
\includegraphics[width=8cm]{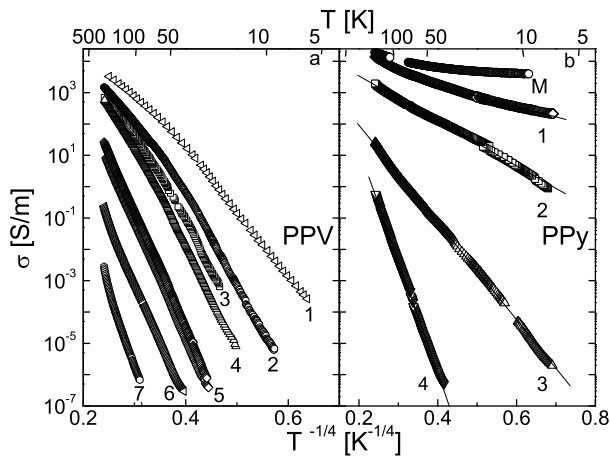}
\end{center}
\noindent{\caption{$\sigma(T)$ versus $T^{-1/4}$ for FeCl$_3$
doped PPV (a) and PPy-PF$_6$ (b) on a logarithmic - linear scale.
At low $T$ $\sigma(T) = \sigma_0 \exp [-(T_0/T)^{1/4}]$; the lines
are fits used to determine $T_0$. (a) The doping levels per
monomer (per nm$^3$) of the samples 1 to 7 are resp. 0.33 (0.69),
0.17 (0.36), 0.10 (0.21), 0.08 (0.17), 0.06 (0.13), 0.03 (0.06),
and 0.02 (0.04).  (b) The doping levels per monomer (per nm$^3$)
of the metallic sample M and the insulating samples 1 to 4 are
0.23 (1.82), 0.16 (1.22), 0.14 (1.14), 0.075 (0.57), and 0.065
(0.51).} \label{PPVPPysigmavsTq}}
\end{figure}

\begin{figure}[htb]
\begin{center}
 \includegraphics[width=6cm]{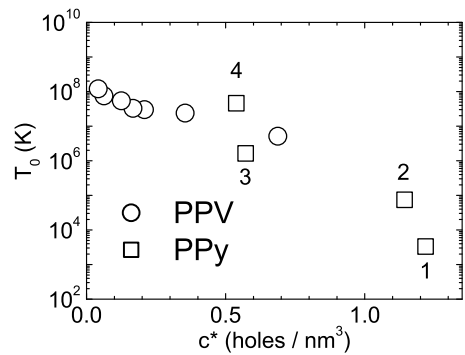}
\end{center}
\noindent{\caption{(a) $T_0$ vs $c^*$ for both FeCl$_3$ doped PPV
(see Ref.~\onlinecite{Martens03}) and PF$_6$ doped PPy. $T_0$ is
determined from the $T^{-1/4}$-dependence of $\log \sigma$, see
Fig.~\ref{PPVPPysigmavsTq}.} \label{gc}}
\end{figure}

{\it Conductivity}. The conductivity $\sigma$  versus $T^{-1/4}$
at various doping levels is shown in Fig.~\ref{PPVPPysigmavsTq} in
logarithmic-linear scale. In both systems the $T$ dependence of
$\sigma$ is quite sensitive to $c$. The most noticeable difference
among PPV and PPy is that $\sigma$ of PPy-PF$_6$ for $c
> 0.16$ follows a real metallic $T$ dependence (large finite
$\sigma$ as $T \rightarrow 0$~K), whereas even in fully doped
OC$_1$C$_{10}$-PPV $\sigma$ still decreases by several orders of
magnitude with $T$. Furthermore in both systems the equation
$\sigma(T) = \sigma_0 \exp [-(T_0/T)^{1/4}]$, expected for
three-dimensional (3-D) variable range hopping (VRH),
\cite{Mott69} fits the data quite well for almost all values of
doping, especially at low temperatures. In the usual analysis
$T_0$ is connected to the density of states $g$ via $k_{\rm
B}T_0(c) \sim 20{\alpha ^3}/{g(E)}$.\cite{Bottger85} The parameter
$\alpha^{-1}$ characterizes the decay of the squared wave function
away from the localization site and equals $0.2
-0.4$~nm.\cite{Bottger85} For doped PPV, the $T_0$ method gives
reliable results for the DOS in the VRH regime at low temperatures
(around 1 state per eV and nm$^3$ in agreement with the EGT data).
For PPV the analysis could be extended by taking into account that
at higher doping levels the size of the delocalized regions
increases. \cite{Martens03} However, for the two highest doped
samples of PPy the values for the DOS determined from $T_0$
($10^2$ - $10^3$ states per eV and nm$^3$) are orders of magnitude
higher (note the logarithmic vertical scale in Fig.~\ref{gc}) than
the ones determined for PPV  or measured by the EGT method. Even
by allowing a growing size of the delocalized
region,\cite{Martens03} no reasonable $g$-values could be
obtained. Apparently the particular character of the disorder in
the polymeric material close to the metal insulator transition
(see Refs. \onlinecite{Kaiser01,Prigodin04,Romijn03,Martens04}),
asks for a more sophisticated analysis of the $T_0$ parameter.

\section{Conclusions}

The DOS per monomer volume as a function of energy, at very
precise values of doping levels, in both OC$_1$C$_{10}$-PPV doped
with FeCl$_3$ and PF$_6$, and PF$_6$-doped PPy has been
determined. An almost linear increase in DOS vs $c^*$ has been
observed in both systems from the EGT data. For PF$_6$ doped PPy
at high $c^*$ the DOS per monomer volume is higher and states
closer to the center of the band can be populated, which
eventually can make the polymer metallic (other parameters like
the inter-chain transfer integrals remain of course essential in
charge transport). This study has also shown that while for doped
PPV interpretation of the data within a VRH picture works well,
for highly doped PF$_6$-PPy such an interpretation might lead to
too high estimates of the density of states.

\begin{acknowledgments}
We acknowledge fruitful discussions with Reinder Coehoorn and
Hubert Martens (Philips Research), Frank Pasveer and Thijs Michels
(Technical University of Eindhoven and Dutch Polymer Institute).
Arjan Houtepen (University of Utrecht) helped us performing the
measurements with the electrochemical transistor and Oleg Bakharev
(Leiden University) with the NMR spectrometer. This work forms
part of the research program of the Dutch Polymer Institute (DPI),
project DPI274.
\end{acknowledgments}

\appendix

\section{The doping level}

\begin{figure}[htp]
\begin{center}
\includegraphics[width=7.5cm]{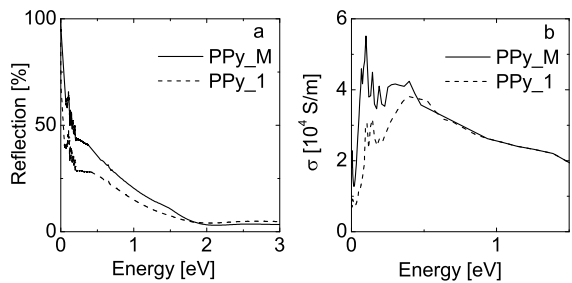}
\end{center}
\caption{(a) IR and UV/VIS reflection of two PF$_6$ doped PPy
samples at 300~K. The overall features are similar. (b) The real
part of the conductivity $\sigma(\omega)$ derived from (a)
according to the procedure in the text. Via the optical sum rule
the real part of $\sigma(\omega)$ or the imaginary component of
the dielectric constant $\epsilon(\omega)$ are related to the
number of carriers. \label{reflection}}
\end{figure}

The PF$_6$ doping levels were determined by use of the optical sum
rule and NMR. Below we explain why we preferred the outcome of the
NMR analysis.

{\it Optical sum rule}. Romijn {\it et al.} used reflection data
in the range 5 meV - 3.5 eV together with the boundary conditions
set by phase sensitive sub-THz spectroscopy to calculate the phase
via the Kramers-Kroning relation ($\theta(\omega_0) = \omega_0 / 2
\pi \int_{0}^{\infty} ln[R(\omega) / R(\omega_0)] / [\omega_0^2 -
\omega^2]d\omega$).\cite{Romijn03} The reflection amplitude and
phase give the real and imaginary parts of the dielectric
constant, see Fig.~\ref{reflection}, where the imaginary component
of the complex relative dielectric constant  $\epsilon_2(\omega)$
(or the real part of the conductivity $\sigma(\omega)=\omega
\epsilon_0 \epsilon_2(\omega)$ with $\epsilon_0$ the vacuum
dielectric constant) is related to the number of carriers via the
sum rule:\cite{Lee95,Romijn03}
\begin{equation}\label{sum}
\frac{N_h(E)}{m^*}=\frac{2 \epsilon_0}{\pi e^2}\int_{0}^{E}\omega
\epsilon_2(\omega)d\omega.
\end{equation}
In this way the ratio $N_h(E)/m^*$ was determined with $N_h(E)$
the number of carriers per m$^3$ and $m^*$ their effective mass.
By making an additional assumption about the effective mass, the
number of carriers was estimated. For $m^*$ equal to the free
electron mass, the number of carriers for PPy\_M found by Romijn
{\it et al.} was about 3 holes/nm$^3$.\cite{Romijn03}

We collected reflection data on PPy samples with very different
room temperature dc conductivities. The outcome of the sum-rule is
somewhat arbitrary, because at energies of ($\sim$~3 eV) intraband
excitations start playing a role as well. \cite{Lee95,Lee04} By
integrating the conductivity up to 3.2 eV, the results show that
in PPy\_4 (notation as in Fig~\ref{PPVPPysigmavsTq}) a carrier
density of 2~holes/nm$^3$ is present while for PPy\_1, see also
Fig.~\ref{reflection}, the carrier density equals 3~holes/nm$^3$;
hence the values of carrier densities in all measured samples are
rather close, though their $\sigma(T)$'s are widely different.

\begin{figure}[htb]
\begin{center}
 \includegraphics[width=5cm]{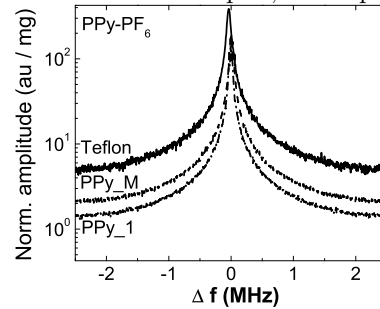}
\end{center}
\noindent{\caption{$^{19}$F signal normalized to the sample mass
(background subtracted) as a function of frequency difference with
the central frequency 376.302 MHz. Shown are Fourier transforms of
free induction decays for teflon, the metallic sample PPy\_M and
sample PPy\_1 (other samples are omitted for clarity).}
\label{NMR}}
\end{figure}

{\it NMR.} A more accurate way to measure the amount of doping is
by nuclear magnetic resonance experiment (NMR). None of the atomic
species present in the dopant (PF$_6$) are contained in PPy.
Because for each P atom there are 6 F atoms and F has spin $I=1/2$
with a very large nuclear magnetic moment, we monitored the F
atoms in 9.4~T at a frequency of 376.302 MHz via the Free
Induction Decay (FID). In Fig.~\ref{NMR}, the signals normalized
to the sample mass of two samples and a teflon (building block
${\rm C_2F_4}$) reference are plotted. The F intensity for each of
the samples is obtained by integration of the signal. The
similarity in line shapes of all F-lines allowed the integration
to be cut-off at the border of the figure without affecting the
intensity ratios.

Because the signal intensity $I^S$ is proportional to the number
of F atoms in the samples, the doping levels per monomer
$c=n_{PF_6}/n_{PPy}$ can be easily determined:
\begin{equation}\label{eq}
  c=[\frac{m_{\rm PPy}}{m_{\rm PF_6}}] \times [{\frac{6 M^S}{4 M^T}\frac{I^T}{I^S}\frac{m_{\rm C_2F_4}}{m_{\rm
  PF_6}}-1}]^{-1},
\end{equation}
where the PPy mass equals the monomer mass ($m_{\rm PPy}$) times
the number of monomers ($n_{\rm PPy}$), the PF$_6$ mass the number
of ions ($n_{\rm PF_6}$) multiplied with the ion mass ($m_{\rm
PF_6}$), and the sample mass
$M^S=m_{PPY}n_{PPy}+m_{PF_6}n_{PF_6}$. $M^T$ and $I^T$ denote
respectively the teflon mass and signal intensity. For samples M,
1 and 4, the determined doping concentrations were respectively
0.23 $\pm$ 0.2, 0.16 $\pm$ 0.02 and 0.065 $\pm$ 0.01 in units of
holes/monomer. From the NMR analysis the insulating sample nr. 4
appears to be almost three times lower doped than the metallic
sample M, which is more realistic than the values obtained from
the optical sum rule.

\end{document}